\documentclass[apj]{emulateapj}

\shorttitle{Temporal evolution of a white-light flare}
\shortauthors{Kawate et al.}

\begin{document}

\title{Temporal evolution and spatial distribution of white-light flare kernels  in a solar flare}

\author{T. Kawate\altaffilmark{1}}
\affil{Institute of Space and Astronautical Science, Japan Aerospace Exploration Agency, 3-1-1 Yoshinodai, Chuo-ku, Sagamihara, Kanagawa, 252-5210 JAPAN}
\email{kawate@solar.isas.jaxa.jp}

\author{T. T. Ishii, Y. Nakatani, K. Ichimoto, A. Asai\altaffilmark{2}}
\affil{Kwasan and Hida Observatories, Kyoto University, Kurabashira, Kamitakara-cho, Takayama, Gifu 506-1314 JAPAN}

\author{S. Morita\altaffilmark{3}}
\affil{Solar Observatory, National Astronomical Observatory of Japan, 2-21-1 Osawa, Mitaka, Tokyo 181-8588 JAPAN}

\and

\author{S. Masuda\altaffilmark{4}}
\affil{Institute for Space-Earth Environmental Research, Nagoya University, Furo-cho, Chikusa-ku, Nagoya 464-8601 JAPAN}

\altaffiltext{1}{Institute of Space and Astronautical Science, Japan Aerospace Exploration Agency}
\altaffiltext{2}{Kwasan and Hida Observatories, Kyoto University}
\altaffiltext{3}{Solar Observatory, National Astronomical Observatory of Japan}
\altaffiltext{4}{Institute for Space-Earth Environmental Research, Nagoya University}

\begin{abstract}
On 2011 September 6, we observed an X2.1-class flare in continuum and H$\alpha$ with a frame rate of about 30~Hz. After processing images of the event by using a speckle-masking image reconstruction, we identified white-light (WL) flare ribbons on opposite sides of the magnetic neutral line. 
We derive the lightcurve decay times of the WL flare kernels at each resolution element by assuming that the kernels consist of one or two components that decay exponentially, starting from the peak time.
As a result, 42\% of the pixels have two decay-time components with average decay times of 15.6 and 587 s, whereas the average decay time is 254 s for WL kernels with only one decay-time component. 
The peak intensities of the shorter decay-time component exhibit good spatial correlation with the WL intensity, whereas the peak intensities of the long decay-time components tend to be larger in the early phase of the flare at the inner part of the flare ribbons, close to the magnetic neutral line.
The average intensity of the longer decay-time components is 1.78 times higher than that of the shorter decay-time components.
 If the shorter decay time is determined by either the chromospheric cooling time or the nonthermal ionization timescale and the longer decay time is attributed to the coronal cooling time, this result suggests that WL sources from both regions appear in 42\% of the WL kernels and that WL emission of the coronal origin is sometimes stronger than that of chromospheric origin.
\end{abstract}

\keywords{Sun: flares, Sun: chromosphere, magnetic reconnection, radiation mechanisms: nonthermal, radiation mechanisms: thermal}

\section{Introduction}

Since Carrington and Hodgson first observed a white-light (WL) flare in 1859 \citep{carr59,hodg59}, WL sources has been observed in intense flares. The typical size of WL sources is 2--3~arcsec and the time duration ranges from several tens of seconds \citep{neid89} to about 10 min \citep{hiei82}. WL sources mainly appears in the impulsive phase of energetic flares, and it has a good temporal and spatial correlations with hard x-ray emission (HXR). However, transporting energy via nonthermal electrons into the photosphere where continuum emission originates in the standard solar atmosphere is difficult \citep{naji70}. Therefore, the emission mechanisms and the source regions of WL flares are still being debated.

\cite{mach89} proposed a model of the formation of WL flares: First, nonthermal electrons directly heat the chromosphere. 
Second, the heated plasmas in the chromosphere emit Balmer continuum photons. 
These photons penetrate downward, raising the temperature of the temperature-minimum region and then exciting the H$^-$ continuum. 
Third, extreme ultraviolet (EUV) emission in the corona penetrates and excites the H$^-$ continuum in the temperature minimum region. 
\citeauthor{mach89} suggest that these processes are coupled in WL flare kernels.
Following the study of \cite{mach89}, these processes were examined numerically by \cite{hawl92}, who compared observed stellar spectra with calculated spectra. 
They found differences in the color temperature and in the magnitude of the Balmer jump, and concluded that the observed continuum forms because of irradiation of intense UV to EUV line emission from the upper chromosphere. 
However, based on a radiative hydrodynamics simulation, \cite{allr05} concluded that EUV irradiation contributes less than 10{\%} of the heating and that irradiation by the Balmer continuum produces more intense WL emission.
These mechanisms for heating (i.e., the sources of EUV, Balmer continuum and HXR) are located at different heights, which stimulated extensive discussion of direct imaging of limb WL flares (e.g. \citealp{batt11,mart12}) and temporal evolution of WL emission compared with that of HXR emission \citep{kane85,huds92,matt03}.
If the dominant mechanism producing WL emission is the recombination of {\sc H~ii} produced by direct heating by nonthermal electrons, the temporal evolution of  WL emission should be determined by both the HXR lightcurve and the recombination timescale of {\sc H~ii}, which is $(4\times10^{-13} N_e)^{-1}$ s \citep{badn06}, where $N_e$ is the electron density in units of cm$^{-3}$.
On the other hand, if the origin of WL emission is radiative heating by the Balmer continuum from the chromosphere or by EUV from the corona, the decay time of WL emission will be comparable to the cooling time of these sources.
\cite{xu06} reported two distinguishable decay times in infrared continuum emission from flare kernels: one of about 30~s and the other of several minutes. 
They concluded that these decay times correspond to cooling times in the chromosphere and corona, respectively, and that the continuum emissions with shorter and longer decay times originate in the different layers. 
From the spatial features, \cite{neid93} confirmed that WL flare ribbons have a bright core in the inner region and a diffuse halo in the outer region. \cite{isob07} also reported the presence of core and halo emissions in a WL flare, and, by assuming that the core is formed by direct heating and that the halo is formed by back-irradiation, they concluded that the halo emission is about 100~km deep or less with respect to the height of the source of radiative back-warming.

One of the difficulties in understanding the mechanisms of WL flares is that they are rare and usually small in size, short in time, and low in contrast; in other words, observing WL flares is difficult. 
Recent observations with high spatial resolution provide more opportunities to detect WL flares. They disclosed that WL emission is sometimes observed even in C-class flares \citep{jess08,wang09}. 
However, the number of observations with high temporal resolution with a cadence better than 1~s is quite limited. \cite{neid93} observed a WL flare in continuum at 5000~\AA \ and H$\alpha$ wings with 0.5~s cadence. The flare was simultaneously observed in HXR by the Hard X-Ray Burst Spectrometer onboard the Solar Maximum Mission satellite with a temporal resolution of 0.128~s but without spatial resolution.
They found that HXR, the H$\alpha$ wing, and WL rise rapidly with about 2.5~s time delay from one to the next, which led them to suggest that this time delay comes either from the ionization timescale of the chromosphere by nonthermal heating or from the formation of a chromospheric condensation.

Since the different excitation mechanisms of WL emission will become manifest on different spatial and temporal scales, observations with both high temporal and spatial resolutions are crucial to understand the excitation process of WL emission. 
In this paper, we present our observations of a WL flare resolved both spatially and temporally. 
These high-resolution observations allow us to examine the decay times in detail at each position in flare ribbons.
Our aim is to examine the spatial distribution of WL kernels and the statistics of decay time in the WL flare ribbons to understand the major cause of WL emission.
This paper consists of the following sections: In Section \ref{sec:obs}, we describe properties of the flare observed in this work and our observation system. In Section \ref{sec:ana}, we show the analysis method and the results. In Section \ref{sec:sum}, we summarize the results and discuss the implication of different timescales observed in the WL flare kernels. Finally in Section \ref{sec:con}, we present our conclusions.

\section{Observations} \label{sec:obs}

\subsection{Overview of a WL flare}
\begin{figure}
\epsscale{1}\plotone{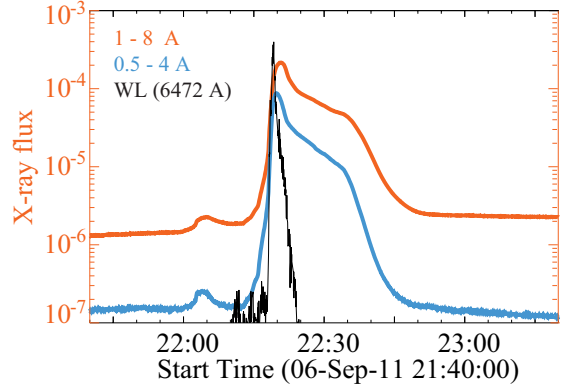}
\caption{GOES soft x-ray fluxes and total WL flux. Red thick, blue thick, and black thin lines show 1 to 8~\AA, 0.5 to 4~\AA, and WL flux in our observation, respectively. The soft x-ray fluxes are in W{\,}m$^{-2}$ unit while WL flux is in arbitrary units.\label{fig:goes}}
\end{figure}

The event we discuss here was an X2.1 class flare that occurred on September 6, 2011 in NOAA11283. 
The soft x-ray flux observed by the Geostationary Operational Environmental Satellite (GOES) started to increase at 22:12~UT, and reached its maximum at 22:20~UT. 
Figure~\ref{fig:goes} shows GOES lightcurves during the entire flare. 
When the flare occurred, the active region was located at N14W18 (i.e., close to the disk center). The flare was observed at different wavelengths by multiple instruments such as the Helioseismic and Magnetic Imager (HMI; \citealp{sche12}) on board the Solar Dynamics Observatory (SDO; \citealp{pesn12}), the Atmospheric Imaging Assembly (AIA; \citealp{leme12}) on board SDO, and the Reuven Ramaty High Energy Solar Spectroscopic Imager (RHESSI; \citealp{lin02}). 
Many papers discuss the flare and the active region. 
For instance, \citet{xu14} analyzed HMI and RHESSI data from this flare and found a two-ribbon structure in HMI continuum images. They confirmed good spatial and temporal correlations between WL and HXR emissions. \citet{jian13} compared the observed coronal magnetic structure by HMI and AIA with the magnetic flux rope by magnetohydrodynamics simulation, and argued that the reconnection at the null cuts likely triggers the torus instability of the flux rope, which result in an eruption.
\citet{roma15} studied HMI observations, and found that the magnetic helicity accumulated in the corona increases monotonically, which appears as high shear and a dip angle in a magnetic field leading to the recurrent flares.

\begin{figure*}
\epsscale{1}\plotone{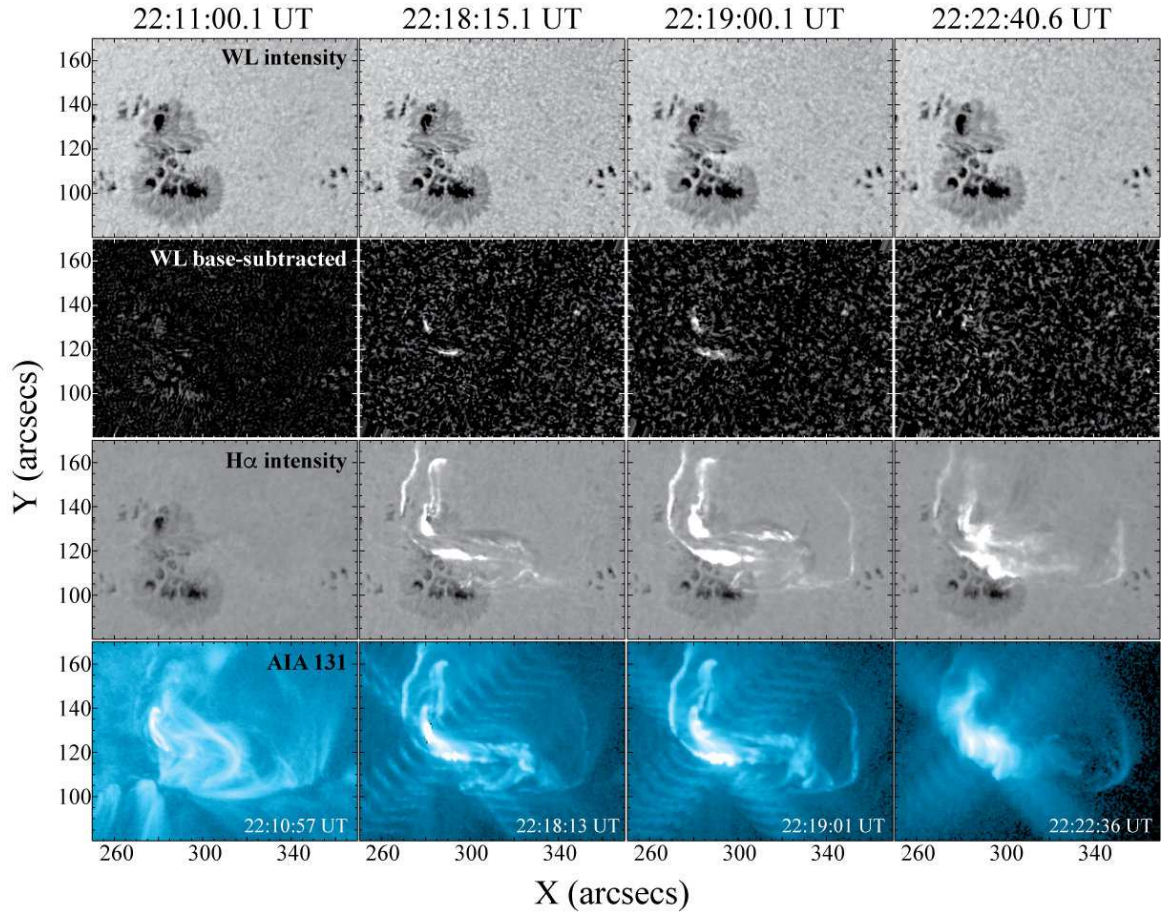}
\caption{Time series of intensity maps of WL, H$\alpha$, AIA 131~\AA , and base-subtracted images of WL.  The time of each set of FISCH images is shown above the panels, and the time of each AIA image is shown at the right-bottom of each image. \label{fig:wlimg}}
\end{figure*}

\subsection{FISCH observation}
We observed the flare by using the Flare Imaging System in the Continuum and H-alpha instrument (FISCH; \citealp{ishi13}) installed in the Solar Magnetic Activity Research Telescope (SMART; \citealp{ueno04}) at Hida Observatory, Kyoto University. The observation system consists of a telescope with a diameter of 25 cm and high-speed cameras with a field of view of 344$''$ $\times$ 258$''$ and a pixel scale of 0.214$''$ pixel$^{-1}$. The time cadence was 30 to 33 frames per second. The observing wavelengths were H$\alpha$  (6563 \AA ) and the red continuum at 6472 \AA \ with bandwidths of 3 and 10 \AA , respectively.
\citet{ishi13} reported the FISCH data of the WL flare as a first result of the instrument, and showed the appearance of a very small and impulsive brightening in WL with a size of 2$''$ $\times$ 3$''$ and duration of about 25 s. By applying the speckle-masking image reconstruction \citep{ichi14} to images taken successively at 1.5 s intervals, we obtained nearly diffraction-limited images from the 25 cm telescope (i.e., a spatial resolution of about 0.65$''$), with a loss of temporal resolution.
 In this paper, we examine the spatial and temporal evolutions of WL emissions observed by FISCH. We concentrate on the data between 22:10:00 and 22:24:40~UT.

The flare was observed during the initial phase of the FISCH operation, and unfortunately several immature configurations appear in the observation described as follows:
Because the data-transfer system was not sufficiently powerful for processing 30-fps images continuously, occasional data gaps occurred with durations less than 20~s.
Synchronization of two cameras was not exact in this observation, and the time difference between the acquisitions of continuum and H$\alpha$ images was at most about 15~ms at maximum.
 The absolute time of the FISCH data had an uncertainty of up to 10~s, although the accuracy of the time stamps between the two cameras was less than 1~ms. 

\subsection{Postprocessing}

After dark subtraction, flat-field correction, and the speckle-masking image reconstruction, the acquisition times of the continuum and H$\alpha$ images are the same. We aligned the position and de-stretched to remove time-varying image deformation from the series of the reconstructed images. 
Figure~\ref{fig:wlimg} shows examples of intensity maps and base-subtracted images after these procedures.
The method of base subtraction is as follows: The dynamics of the photosphere may affect base subtraction, and the intensity fluctuation of granules due to their time evolution is non-negligible for analyzing WL flares even over a 10~s period.  
We derive low-pass-filtered images in the time domain at the critical frequency of 3.3 mHz (i.e., a period of 5~min) from the series of images taken between 22:10 and 22:15~UT, and subtract the low-pass-filtered images from each image during the flare.
As for the H$\alpha$ images, the photospheric patterns have negligible effects on the base-subtracted images, so we determine the background intensities from the nonflaring image taken at 22:10:00~UT, which is before the onset of the GOES x-ray flux. As a reference of the coronal magnetic features, we also display in Figure~\ref{fig:wlimg} the AIA 131~\AA\ images.
This emission is primary from Fe XXI formed around 10~MK \citep{leme12} in flaring regions.
In the base-subtracted images, we see  two major ribbons in both WL and H$\alpha$ connected by flare loops that are visible in the 131~\AA\ image taken at 22:22:36~UT.
These magnetic configurations were also discussed in previous studies of this flare
(see, e.g., \citealp{jian13}, \citealp{liu14}).

\section{Analyses and Results}\label{sec:ana}
\subsection{Extraction of WL flare kernels}

\begin{figure}
\epsscale{1}\plotone{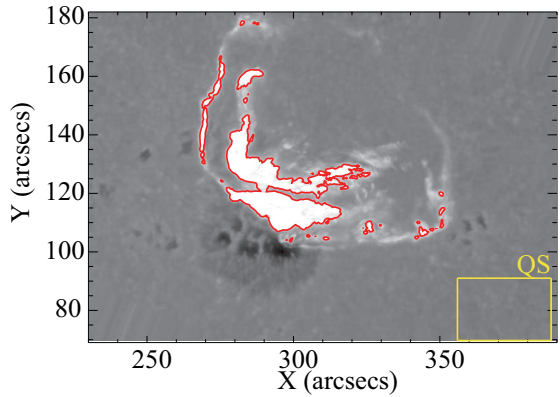}
\caption{Map of $I_{max}$(H$\alpha$). The region of $I_{max}$(H$\alpha$)/$I_{QS}$(H$\alpha$)$>1.5$ is shown in the red contours. The quiet region that we used for normalization is marked by the yellow rectangle.
 \label{fig:ribbon}}
\end{figure}

To begin, we extract the pixels of WL flare kernels in the flaring region. 
During a flare, intensities at H$\alpha$ flare ribbons are enhanced by several times above the quiet-sun (QS) intensity $I_{QS}$(H$\alpha$)~\citep{kita90,fang03}. 
Since we observed the flare with the pass band of 3~\AA , which is much broader than ordinal H$\alpha$ filters used in flare observations,  this contrast may decrease. 
We set a threshold of $1.5 I_{QS}$(H$\alpha$), and pick up pixels at which the intensities exceed the threshold at least once between 22:10 and 22:24~UT. 
We determine $I_{QS}$(H$\alpha$) for each frame by averaging intensities over the region spanning 356$^{\prime\prime}$ to 388$^{\prime\prime}$ along the east-west axis and 69$''$ to 91$''$ along the north-south axis in the solar coordinate system. 
The area where H$\alpha$ ribbons are detected is 15680~pixels, which is 720~arcsec$^2$ in our observation. 
Figure~\ref{fig:ribbon} shows the detected H$\alpha$ flare ribbons and the quiet region used for the intensity reference.

We now estimate the error in the measurement of the intensity enhancement of WL emission. 
Standard deviations of the base-subtracted intensity in space may increase over time because of the evolution of granules and sunspots, since we define the background intensity as the preflare intensity. 
The standard deviation of base-subtracted continuum intensities is 2.3\% of $I_{QS}$(WL) before the flare between 22:10:00 and 22:12:00~UT in the area of H$\alpha$ flare ribbons, whereas it is 4.4\% of $I_{QS}$(WL) after the GOES peak flux between 22:22:40 and 22:24:40~UT, where $I_{QS}$(WL) is the QS intensity of WL determined in the same region as $I_{QS}$(H$\alpha$). 
We evaluate the error in the measurement of the WL intensity by averaging these standard deviations before the flare and after the peak (i.e.,  it is 3.4\% in one sigma).

We choose pixels of WL kernels under the following two criteria: 
(1) The normalized maximum intensity enhancement of WL, $[ I_{max}(WL)-I_{BG}$(WL)]/$I_{QS}$(WL), is larger than  3.4\%, where $I_{max}$ is the maximum intensity and $I_{BG}$ is the preflare intensity at each pixel, which is also used to derive base-subtracted images discussed above.
(2) When the WL intensity reaches the maximum in a pixel, the intensity of H$\alpha$ at this pixel exceeds $1.5 I_{QS}$(H$\alpha$). As a results, 4546 pixels (210~arcsec$^2$) are extracted. 
The total flux of base-subtracted WL emission in these pixels is displayed in Figure~\ref{fig:goes}.
In the following subsection, we examine the lightcurve at each pixel.

\subsection{Temporal evolution of each WL flare kernels}

\begin{figure}
\epsscale{1}\plotone{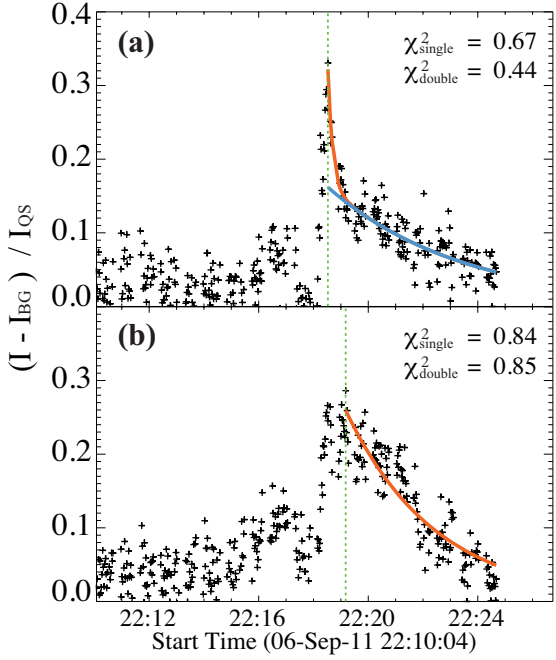}
\caption{(a) An example of a lightcurve for a single pixel and its fit that identifies it as having two components. 
(b)  An example of a lightcurve for a single pixel and its fit that identifies it as having a single component.
Red and blue curves in panels (a) and (b) are the exponential curve fits to the decaying components with time constants $t_{decay1}$ and $t_{decay2}$, respectively. The vertical dotted green lines show the peak of the WL enhancement.
Reduced $\chi ^2$ of the fittings to single component ($\chi ^2_{single}$) and double components ($\chi ^2_{double}$) are shown at the top-right side in each panel.
\label{fig:wllc}}
\end{figure}

\begin{figure}
\epsscale{1}\plotone{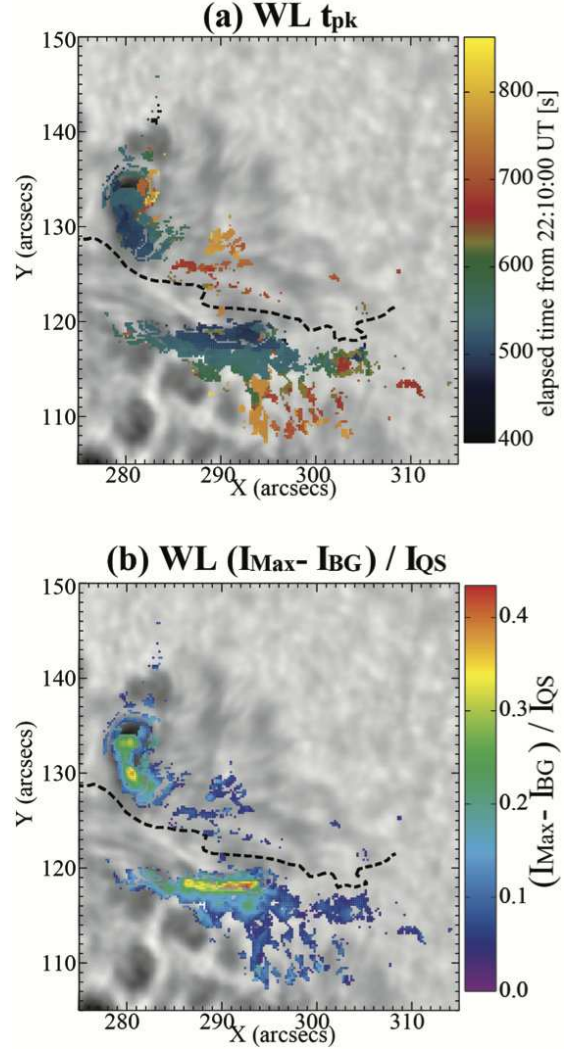}
\caption{Maps of temporal evolution of WL emission. 
 (a) Peak time $t_{pk}$ of WL (colored dots). 
 (b) Peak intensity of WL $[I_{max}$(WL)$ - I_{BG}$(WL)]/$I_{QS}$(WL) (colored dots). 
The greyscale background is the preflare WL intensity map at 22:10:00~UT. 
Thick, black dotted curve shows the magnetic neutral line.
\label{fig:proptot}}
\end{figure}

\begin{figure*}
\epsscale{1}\plotone{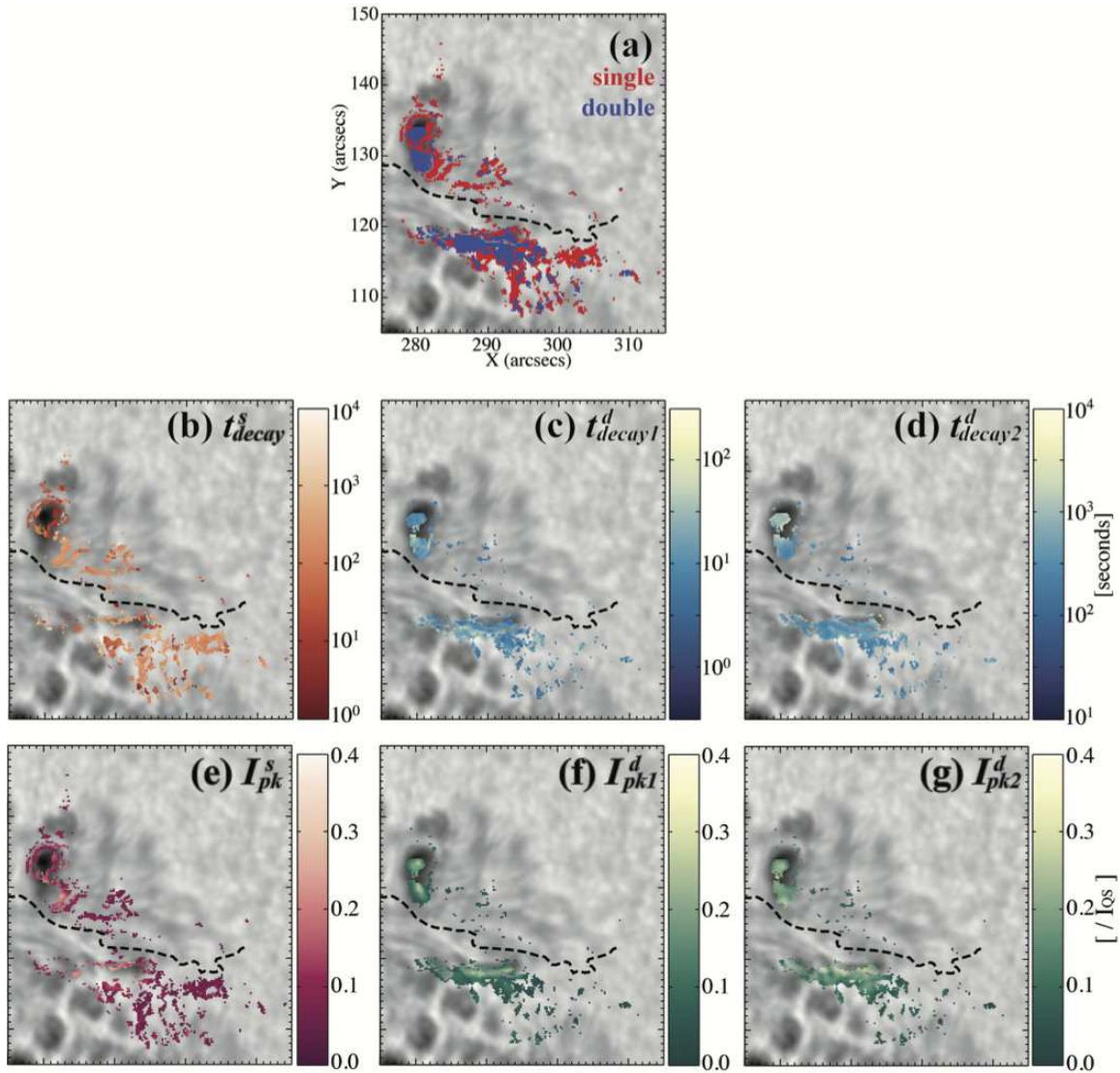}
\caption{Maps of intensity and temporal evolutions of WL emission. 
(a) Pixels that are better fit to single component (red) and to double components (blue).
(b) Decay time of WL of single component $t_{decay}^s$ (colored dots).
(c) Decay time of WL of shorter timescales in double components, $t_{decay1}^d$ (colored dots). 
(d) Decay time of WL of longer timescales in double components, $t_{decay2}^d$ (colored dots).
(e) Peak intensity $I_{pk}^s$ of single component (colored dots).
(f) Peak intensity $I_{pk1}^d$ of shorter decay time in double components (colored dots).
(g) Peak intensity $I_{pk2}^d$ of longer decay time in double components (colored dots).
The greyscale background is the WL intensity map in the preflare at 22:10:00~UT. 
Thick, black dotted curve shows the magnetic neutral line.
\label{fig:proptim}}
\end{figure*}

In the next step, we examine the lightcurve of WL flare kernels, and derive the peak time $t_{pk}$ and decay time $t_{decay}$ at  each position. 
Since there are occasional data gaps, we discard the data at which the $t_{pk}$ detected is within three frames from the edge of the data gap. 
Regarding $t_{decay}$, if WL emission is produced by multiple sources, the timescale of the decaying feature may also have multiple components, as has been discussed by various researchers (see, e.g., \citealp{huds92}, \citealp{xu06}). 
Here, we assume one or two components, each of which decays exponentially; in other words, the light curve in the decay phase is represented by either of the following two equations:
\begin{eqnarray}
\frac{I(t)-I_{BG}}{I_{QS}} &=& I_{pk1}^d \exp(-\frac{t}{t_{decay1}^d})+I_{pk2}^d \exp(-\frac{t}{t_{decay2}^d}) \\
\frac{I(t)-I_{BG}}{I_{QS}} &=& I_{pk}^s \exp(-\frac{t}{t_{decay}^s})  ,
\end{eqnarray}
where $t$ is time elapsed since $t_{pk}$, $I_{pk}^s$, $I_{pk1}^d$, and $I_{pk2}^d$ are the intensity of each component at the peak time, and $t_{decay}^s$, $t_{decay1}^d$, and $t_{decay2}^d$ are the decay time of each component, with $t_{decay1}^d<t_{decay2}^d$. 
 We denote the parameters of the single decay-time component by $X^s$ and those of the double decay-time component by $X^d$.
 We fit the lightcurve at each pixel with this function in the time interval from $t_{pk}$ to 22:24:40~UT. Since the time resolution of our observation was 1.5~s, we set the lower limit of $t_{decay}$ to be the inverse Nyquist frequency of the time resolution (i.e., 3.0~s).
As for the upper limit of $t_{decay}$, because the GOES flux returned back to the background level at 23:00~UT, the duration of WL emission should be shorter than for the soft x-ray flare (22:12--23:00~UT).
Thus, we set the upper limit of $t_{decay}$ to be 3.0 $\times 10^3$~s. 
If either $t_{decay1}^d$ or $t_{decay2}^d$ reach these limits or reduced $\chi ^2$ of the fitting by equation (2) (single component) is smaller than that by equation (1) (double components), we assume that the lightcurve has only a single decaying component. 
If after the fitting all $t_{decay}$ are beyond these limits or both reduced $\chi ^{2}$ of single- and double-component fittings are greater than 3, then we discard the lightcurve. 
We display examples of lightcurves and the results of fitting them to double and single components in Figures~\ref{fig:wllc}(a) and \ref{fig:wllc}(b), respectively.
As a result, two components are detected in 1490 lightcurves, and a single component is detected in 2031 lightcurves, which correspond to 69 and 94 arcsec$^2$, respectively. 
This analysis suggests that 42\% [=1490/(1490+2031)] of the WL flare kernels have at least two decay-time components. 
We show the spatial distribution of $I_{pk1}+I_{pk2} \equiv I_{max}$ and $t_{pk}$ in Figure~\ref{fig:proptot}. The magnetic neutral line derived from the line-of-sight photospheric magnetic field taken by SDO/HMI is overlaid in each figure.
 We can see that the detected $t_{pk}$ propagates outward from the magnetic neutral line in Figure~\ref{fig:proptot}(a), and the regions close to the magnetic neutral lines are brightest in the ribbons in Figure~\ref{fig:proptot}(b). 
 The spatial distribution of pixels that are better fit to single and double components are shown in Figure~\ref{fig:proptim}(a).
We also plot $t_{decay}^s$, $t_{decay1}^d$, $t_{decay2}^d$, $I_{pk}^s$, $I_{pk1}^d$, and $I_{pk2}^d$ in the same manner in Figure~\ref{fig:proptim}(b) to \ref{fig:proptim}(g) as Figure~\ref{fig:proptot}.
We find no systematic trend in spatial distribution of $t_{decay}$ in either the single or double components in Figures~\ref{fig:proptim}(b), \ref{fig:proptim}(c), and \ref{fig:proptim}(d), whereas the spatial distribution of $I_{pk1}$ and $I_{pk2}$ are similar to the distribution of $I_{max}$ and $t_{pk}$, respectively, in Figures~\ref{fig:proptim}(f) and \ref{fig:proptim}(g).
 The cross-correlation coefficients between $t_{pk}$ and $t_{decay}$ of each component, $t_{pk}$ and $I_{pk}$ of each component, $I_{max}$ and $t_{decay}$ of each component, and $I_{max}$ and $I_{pk}$ of each component are summarized in Table~\ref{tab:cc}. 
The cross-correlation coefficients between $t_{pk}$ and $t_{decay}$ and $I_{max}$ and $t_{decay}$ are between -0.25 and 0.25.
However, $I_{pk1}^d$ correlates strongly with $I_{max}$ with a cross-correlation coefficient of 0.88, whereas the cross-correlation coefficient between $I_{pk2}^d$ and $I_{max}$ is 0.73. This means that the spatial distribution of $I_{pk1}^d$ is particularly similar to that of $I_{max}$. 
The cross-correlation coefficients between $t_{pk}$ and $I_{pk1}^d$ and between $t_{pk}$ and $I_{pk2}^d$ are -0.37 and -0.56, respectively (i.e., the kernels tend to be brighter in the early phase of the flare). 
These results suggest that the morphology of WL emission with the shorter decay-time component is more similar to that of the  WL emission  than that of the longer decay-time component. In other words, they resemble compact and bright WL kernel.
However, since $t_{pk}$ increases and $I_{pk2}^d$ decreases gradually from the inner part to the outer part of the flare ribbons, the WL emission with longer decay time is mainly excited in the inner part of the active region close to the magnetic neutral line.

\begin{table}
\begin{center}
\caption{Cross-correlation coefficients between $t_{pk}$ and $t_{decay}$ of each component, $t_{pk}$ and $I_{pk}$ of each component, $I_{max}$ and $t_{decay}$ of each component, and $I_{max}$ and $I_{pk}$ of each component. \label{tab:cc}}
\begin{tabular}{crrrrrr}
\tableline\tableline
& $t_{decay}^s$ & $t_{decay1}^d$ & $t_{decay2}^d$ & $I_{pk}^s$ & $I_{pk1}^d$ & $I_{pk2}^d$ \\
\tableline
$t_{pk}$ & -0.19 & -0.23 & -0.12 & -0.40 & -0.37 & -0.56 \\
$I_{max}$ & 0.019 & -0.067 & 0.013 & 0.80 & 0.88 & 0.73 \\
\tableline
\end{tabular}
\end{center}
\end{table}

\begin{figure}
\epsscale{1}\plotone{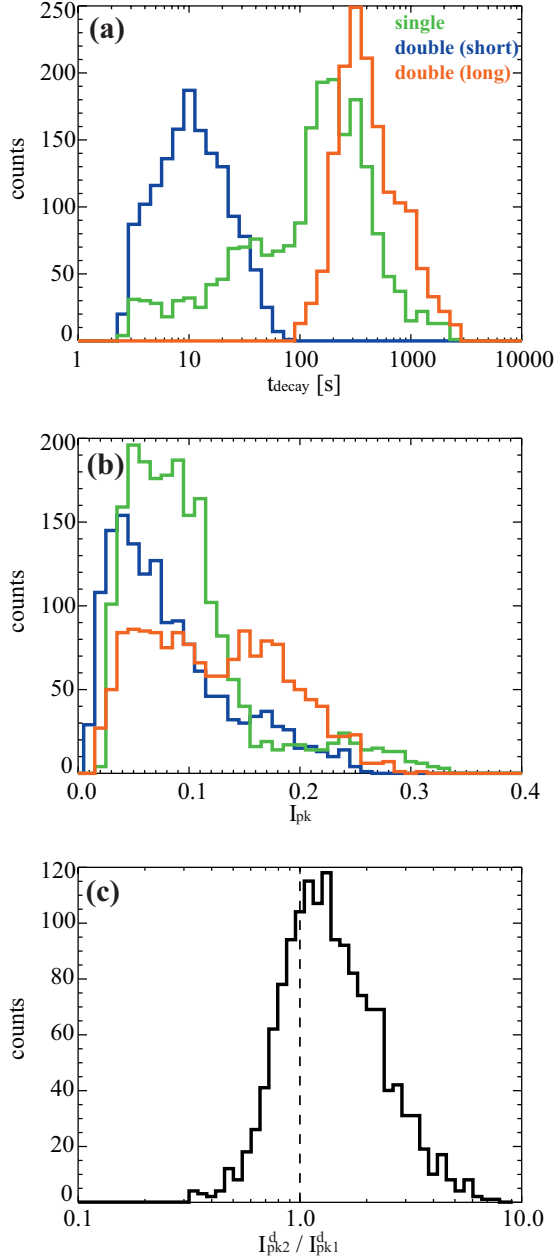}
\caption{
(a) Histograms of decay time $t_{decay}$ of single component (green), shorter decay time in double components (blue), and longer decay time in double components (red)
(b) Histograms of peak intensity $I_{pk}$ of single component (green), shorter decay-time component in double components (blue), and longer decay-time component in double components (red).
(c) Histogram of ratio between peak intensity of shorter decay-time component and peak intensity of longer decay-time component in double components.
 \label{fig:histo}}
\end{figure}

Figures~\ref{fig:histo}(a) and \ref{fig:histo}(b) show histograms of $t_{decay}$ and $I_{pk}$, respectively, for each component.
The average of $t_{decay}^s$ is 254 s with a standard deviation of 316 s, and the averages of $t_{decay1}^d$ and $t_{decay2}^d$ are 15.6 s and 587 s, respectively, with standard deviations of 12.2 s and 460 s, respectively. 
The average of $I_{pk}^s$ is 0.090 with a standard deviation of 0.044, and the averages of $I_{pk1}^d$ and $I_{pk2}^d$ are 0.088 and 0.129, respectively, with standard deviations of 0.055 and 0.063, respectively.
To examine which component is a major fraction of WL emission, we derive the ratio between the peak intensity of the shorter decay-time component and the peak intensity of the longer decay-time component,  $I_{pk2}^d/I_{pk1}^d$, and show the result in the histogram in Figure~\ref{fig:histo}(c). The average of $I_{pk2}^d/I_{pk1}^d$ is 1.78 with a standard deviation of 1.06 (i.e., statistically, the intensity of the longer decay-time component is greater than that of the shorter decay-time component). Of the area that shows double decay-time components, 79\% shows higher intensities for the longer decay-time components than for the shorter decay-time components.
In the histograms, no significant difference appears between the populations of $I_{pk1}^s$ and $I_{pk1}^d$ and between those of $t_{pk1}^s$ and $t_{pk1}^d$.

\section{Summary and Discussions}\label{sec:sum}

We observed WL flare ribbons using SMART/FISCH.
By looking into the lightcurve at each pixel, we derived the peak intensity and the decay times with the assumption that the lightcurve has one or two decaying components. Of the WL flaring lightcurves, 42\% are better fit by the decaying function with two components, and the timescales are 15.6 and 587~s.
The peak intensity enhancements of the lightcurves normalized by the QS intensity are 0.088 and 0.129 for the shorter and longer decay-time components, respectively. As for the lightcurves that are well fit by a single decay component, the decay time is 254~s, and the peak intensity enhancement is 0.090.
The peak intensities of shorter decay-time components correlate strongly with the total peak intensity enhancement of the WL emission, whereas the higher intensities of longer decay-time components tend to appear in the inner part of the flare ribbons, close to the magnetic neutral lines. The average intensity of the longer decay-time components is 1.78 times higher than that of the shorter decay-time component.
%

The typical timescales of the decay are 16 and 590~s, which suggests that both emission components (i.e., one with shorter decay time and one with longer decay time) are excited by the chromospheric and coronal origins, respectively, as discussed in \cite{xu06}. 
According to this reasoning, WL emission of the coronal origin can be stronger than that of chromospheric origin, since the average peak intensity of the longer timescale component is 1.68 times greater than that of the shorter timescale component. 
This result contradicts the results from the one-dimensional numerical study by \cite{allr05}, which found that EUV irradiation contributes to the WL emission much less than does the Balmer continuum formed in the chromosphere.
This may be explained by the size of footpoint and the loop morphology. Nonthermal footpoints are suggested to be highly localized in the H$\alpha$ flare ribbons (see, e.g., \citealp{asai02}). Also, from recent observations, the size of the H$\alpha$ kernel in a flare is 110 to 161~km \citep{jing16}, which is smaller than the spatial sampling size of our observation. 
Thus, the shorter decay-time component in the WL emission may result from the compact brightening in the chromosphere in a small fraction of a single pixel, whereas the longer decay-time component in the WL emission may result from diffuse irradiation around the small kernel. 
The shorter decay-time component is smoothed into one pixel, and the longer decay-time component is accumulated more easily than the shorter decay-time component. 
Thus, the apparent fraction of intensity of coronal origin may increase compared with the one-dimensional atmosphere.
Moreover, the flare loops connected flare kernels in the northern and southern ribbons, and the loops formed successively from inside to outside in the ribbons. Thus, the WL emission from the lower corona may be accumulated in the inner part of the flare structure close to the magnetic neutral line.

The decay time of the single decay-time component is widely distributed from 3.1 to 2700 s with the average value of 250 s.
Thus, both possibilities, i.e., the WL emission originates from the corona and from the chromosphere, arise to explain the origin of single decay-time component. Here, we examine two possibilities separately;
one possibility is that the lightcurve has only one component with a coronal origin. 
In Figure~\ref{fig:proptim}(a), regions that are better fit to single component tend to be located outer in the flare ribbons. If we agree the three-dimensional structure of the coronal irradiation described above, WL emission from the coronal origin should be distributed diffusely around the small kernels. Therefore, WL emission from outer parts in the flare ribbons may only be created by the coronal irradiance. 
This diffuse halo structure located outer in flare ribbons was also discussed in \cite{neid93}.
If the longer decay time in double components is created by the irradiation from all the surrounding pixels, the decay time of WL emission at the edge of the flare ribbons is shorter than those in the inner part of the flare ribbons. This can explain that the histogram of $t_{decay}^s$ peaks at a shorter timescale than that of $t_{decay}^d$ in Figure~\ref{fig:histo}(a).
The other possibility is that the lightcurve has only one component with a chromospheric origin or that the coronal origin of WL kernels is negligible.
The density increases the effect of the EUV emission in the transition region and in the corona; this increase is caused by mass motion during chromospheric evaporation~\citep{anto84}.
Redshift and blueshift of EUV lines have been observed at footpoints in flares \citep{mill09,wata10t}, and the Doppler velocity is smaller if the thermal conduction dominates~\citep{imad15}. Thus, we can assume that the thermal conduction is strong in the WL flare kernels that have a single decay-time component with decay times of up to several tens of seconds.

\section{Conclusion}\label{sec:con}
Observations with high spatial and temporal resolution allowed us to resolve the cooling process of WL flare kernels. 
Of the WL flare kernels, 42\% have two decay-time components in the lightcurve, with typical timescales of 16 and 590~s.
The difference between the timescales may be determined by the position of the heating sources that create the WL emission either by direct heating by nonthermal electrons or by back-irradiation from the chromospheric or coronal source. 
The total flux of the WL emission from coronal irradiance is sometimes the major cause of WL enhancement in the inner part of the flare ribbons because of the accumulation of radiative heating sources; either the coronal irradiance or the chromospheric origin such as the chromospheric irradiance or nonthermal heating is the candidate of the origin of WL emission for the 58\% of WL kernels that show only single decay-time feature in lightcurve.

\acknowledgments
The authors are grateful to the staff of Hida observatory for helping with the observations.
 We also thank Dr. S. Imada, Dr. H. Isobe, Dr. H. Hudson, and anonymous referees for fruitful discussions and comments. 
 This work was carried out by the joint research program of the Solar-Terrestrial Environment Laboratory and the Institute for Space-Earth Environmental Research, Nagoya University, and was supported by MEXT/JSPS KAKENHI Grant No. JP15H05814.
The authors would like to thank Enago (www.enago.jp) for the English language review.

\bibliographystyle{apj}
\bibliography{refbib}



\end{document}